# Solitary modes in nonlocal media with inhomogeneous self-repulsive nonlinearity

Yingji He[1] and Boris A. Malomed[2]

[1]*School of Electronics and Information, Guangdong Polytechnic Normal University, 510665 Guangzhou, China*

[2]*Department of Physical Electronics, School of Electrical Engineering, Faculty of Engineering, Tel Aviv University Tel Aviv 69978, Israel*

**Abstract**

We demonstrate the existence of two species of stable *bright solitons*, fundamental and dipole ones, in one-dimensional *self-defocusing* nonlocal media, with the local value of nonlinearity coefficient having one or several minima and growing at any rate faster than |*x*| at large values of coordinate *x*. The model can be derived for a slab optical waveguide with thermal nonlinearity. The most essential difference from the local counterpart of this system is the competition between two different spatial scales, the one determining the modulation pattern of the nonlinearity coefficient, and the correlation length of the nonlocality. The competition is explicitly exhibited by analytically obtained asymptotic form of generic solutions. Particular exact solutions are found analytically, and full soliton families are constructed in a numerical form. The multi-channel settings, with two or three local minima of the nonlinearity coefficient, are considered here for the first time, for both local and nonlocal models of the present type. States with multiple solitons launched into different channels are stable if the spacing between them exceeds a certain minimum value. A regime of stable Josephson oscillations of solitons between parallel channels is reported too.

**PACS numbers**: 42.65.Tg, 42.65.Sf; 05.45.Yv

# 1. Introduction

Nonlocality is an inherent feature of many settings in optics, plasmas, and Bose-Einstein condensates. Spatial-domain nonlinear dynamics in nonlocal optical media with different response functions, characterized by the respective correlation lengths, were studied in detail theoretically and experimentally [1]. In particular, the nonlocal nonlinearity can support solitons in various forms [2,3], including fundamental ones [4,5], vortices [5-7], and multihump modes [8-12]. In addition to many numerical results, various soliton solutions, including dipoles, double, and periodic solitons, can be found in an analytical form in nonlocal Kerr media with the exponential response function [13]. Interactions between solitons too are strongly affected by the nonlocality [14]. It has been demonstrated that the nonlocal nonlinear response allows for suppression of the modulation instability of plane waves, arrest of the collapse of multidimensional beams, and, generally, fosters the stabilization of solitons in nonlocal nonlinear media [15,16].

In this work, we introduce a one-dimensional model with a spatially modulated strength of repulsive nonlocal interactions, the strength growing from the center towards the periphery. We demonstrate that, rather counter-intuitively, this *self-defocusing* medium readily supports stable *bright* solitons, both fundamental and dipole-mode ones. In the model which includes several local minima of the self-repulsion strength (*channels*), interactions between solitons trapped in adjacent channels, and regimes of periodic switching between them (*Josephson oscillations* of the solitons) are investigated too.

The possibility of supporting stable bright solitons of various types, in the space of any dimension, by means of the spatially growing self-repulsive local nonlinearity, was proposed recently [17-21], but, thus far, this possibility was not implemented in models of nonlocal media. A new essential feature introduced by the nonlocality is the competition between its correlation length and the spatial scale of the nonlinearity modulation. Nonlocal models of this type are considered here for the first time, as well as settings with multiple channels and interactions between them, which were not studied before in the context of local models either. What we report here also appears to be the first example of Josephson oscillations of solitons in a nonlocal system.

The paper is organizes as follows. The model is formulated in Section 2. Particular analytical solutions for fundamental and dipole solitons, along with asymptotic analytical results for generic solutions, are reported in Section 3. Numerical results, demonstrating the existence of the general families of fundamental and dipole solitons (which incorporate the particular exact solutions of both types) and their stability, are collected in Section 4. The same section includes numerical results obtained for multi-soliton states and Josephson oscillations in multi-channel systems. The paper is concluded by Section 5. The derivation of the model for the optical slab waveguide with thermal nonlinearity is recapitulated in Appendix.

## 2. The model

The normalized form of the propagation equation for envelope $u(x,z)$ of the electromagnetic field in the nonlocal nonlinear medium with the spatially growing coefficient of the repulsive self-interaction, $\sigma(x)$, is

$$iu_z + \frac{1}{2}u_{xx} - mu = 0, \qquad (1)$$

$$m - dm_{xx} = \sigma(x)|u|^2, \qquad (2)$$

where $d > 0$ is, as usual, the squared correlation length of the nonlocality, and real field $-m$ is a local perturbation of the refraction index. In terms of optical media with the thermal nonlinearity [1], Eq. (1) implies that heating of the medium by the beam reduces the local refractive index (in particular, because of the thermal expansion, which makes the density of the material smaller). It is also assumed that the intensity of the heating is determined by the local density of an absorptive dopant, $\sigma(x)$, growing at $|x| \to \infty$. It is relevant to stress that, although in the model adopted below $\sigma(x)$ formally diverges at $|x| \to \infty$, this apparently unphysical condition is not really necessary, as the corresponding self-trapped modes are strongly localized [see, e.g., Eqs. (7), (14) and (8), (13), (19) below], hence the spatial growth of density $\sigma(x)$ may terminate at finite $|x|$.

Our analysis demonstrates that the presence of the first term in Eq. (2), which actually defines a finite correlation length of the nonlocal system, is necessary for the existence of solitons. Therefore, in Appendix we recapitulate the derivation of this equation from the heat-conductivity equation for the optical waveguide.

Stationary solutions to Eqs. (1) and (2) are looked for as

$$u(x,z) = \exp(ibz)U(x), \quad m = M(x), \qquad (3)$$

with real functions $U$ and $M$ obeying the following equations:

$$-bU + \frac{1}{2}U'' - MU = 0, \qquad (4)$$

$$M - dM'' = \sigma(x)U^2, \qquad (5)$$

with the prime standing for $d/dx$. As shown below in Section 3, physically relevant localized modes, with a finite total power,

$$P = \int_{-\infty}^{+\infty} U^2(x)dx, \qquad (6)$$

which is the dynamical invariant of Eq. (1), exist if $\sigma(x)$ grows at any rate faster than $|x|$ at $|x| \to \infty$ (in the two-dimensional version of the model, the necessary condition is the growth faster than $r^2$). Although this condition is the same as in the above-mentioned local models [17-20], an essential difference in the structure of the solitons is that, while $U(x)$ must decay fast enough at $|x| \to \infty$, to provide for the convergence of the total power (6), the real field $M(x)$ does not need to decay at $|x| \to \infty$. In fact, it attains a constant value in this limit, see below.

## 3. Exact and asymptotic solutions for localized modes

### 3.1 The exact solution for the fundamental soliton

Examples of *exact* analytical solutions can be obtained for several types of the nonlinearity-modulation function in Eq. (5), $\sigma(x)$. In particular, this is possible for the following modulation profile:

$$\sigma(x) = \sigma_{-2} \cosh^2(ax) + \sigma_0 - \sigma_2 \text{sech}^2(ax), \qquad (7)$$

with $\sigma_{-2} > 0$. The first example represents the fundamental soliton, which is looked for as

$$U(x) = A\,\text{sech}(ax), \quad M(x) = M_0 - M_2 \text{sech}^2(ax). \qquad (8)$$

This is an exact solution provided that parameters of modulation profile (7) are subject to the following constraint:

$$\sigma_0 = \frac{2}{3}\left(1 - \frac{1}{4da^2}\right)\sigma_2, \qquad (9)$$

the respective coefficients of exact solution (8) and propagation constant being

$$A^2 = \frac{6da^4}{\sigma_2}, \quad M_0 = \frac{6\sigma_{-2}}{\sigma_2} da^4, \quad M_2 = a^2, \qquad (10)$$

$$b = \frac{a^2}{2}\left(1 - \frac{12\sigma_{-2}}{\sigma_2} da^2\right). \qquad (11)$$

This solution is meaningful ($A^2 > 0$) if $\sigma_2$ in Eq. (7) is positive, in addition to the above condition of $\sigma_{-2} > 0$, while the sign of $\sigma_0$ is determined by Eq. (9). Further, the modulation profile (7) is relevant for the local absorptive coefficient if condition $\sigma(x) \geq 0$ holds at all $x$. Taking into regard constraint (9), we conclude that the latter condition implies the following restriction:

$$\sigma_{-2} \geq \frac{1}{3}\left(1 + \frac{1}{2da^2}\right)\sigma_2. \qquad (12)$$

It follows from this inequality that the propagation constant $b$, given by Eq. (11), is always negative.

### 3.2 The exact solution for the dipole soliton (DS)

An exact DS solution to Eqs. (4) and (5), with $U(-x) = -U(x)$ and

$M(-x) = M(x)$, is looked for as

$$U(x) = B \sinh(ax) \operatorname{sech}^2(ax), \quad M(x) = M_0 - M_2 \operatorname{sech}^2(ax), \tag{13}$$

cf. Eq. (8). In this case, the modulation function may be taken in a form simpler than expression (7) adopted above for the fundamental soliton:

$$\sigma(x) = \sigma_{-2} \cosh^2(ax) + \sigma_0 \tag{14}$$

(the former term $\sim \sigma_2$ is not necessary for finding the analytical DS solution). The exact solution in the form of Eq. (13) exists if parameters of modulation profile (14) are subject to the following constraint, cf. Eq. (9):

$$\sigma_{-2} = \frac{1}{3}\left(1 + \frac{1}{2da^2}\right)\sigma_0. \tag{15}$$

Coefficients of the exact DS solution are [cf. Eqs. (10) and (11)]:

$$B^2 = \frac{18da^4}{\sigma_0}, \quad M_0 = \frac{18\sigma_{-2}}{\sigma_0}da^4, \quad M_2 = 3a^2, \tag{16}$$

$$b = -\frac{a^2}{2}\left(\frac{1}{2} + 3da^2\right). \tag{17}$$

Note that Eq. (17) again demonstrates that the exact solution has $b < 0$.

As mentioned above, field $M(x)$ does not vanish in the solutions at $|x| \to \infty$; instead, it attains constant values $M_0$, see Eqs. (10) and (16). However, this feature does not contradict the localized character of the fundamental and DS modes, as their power density is localized.

### 3.3 Asymptotic results for generic solutions

The above exact solutions yield isolated solutions at single values of the propagation constant, $b$, given by Eqs. (11) and (17) for the fundamental and dipole modes, severally. As shown below by means of numerical solutions, generic localized solutions form continuous families parameterized by $b$. To include the exact solution (8)-(11) into the numerically found family of fundamental solitons [see Fig. 1(a) below], we take the modulation profile of the nonlinearity coefficient, $\sigma(x)$, in the form of Eq. (7) with $d = \sigma_2 = 1$, $\sigma_0$ fixed as per Eq. (9), and

$$\sigma_{-2} = \frac{1}{3}\left(1 + \frac{1}{2da^2}\right)\sigma_2, \qquad (18)$$

as suggested by Eq. (12). This choice of the parameters implies that $\sigma(x=0) = 0$ and $\sigma(x>0) > 0$.

It is possible to obtain some asymptotic results for the generic solutions. In the limit of $|x| \to \infty$, a straightforward expansion of Eqs. (4), (5) and (7), which makes use of approximation $\sigma(x) \approx (\sigma_{-2}/4)\exp(2a|x|)$, yields the following asymptotic form of the solution for $|x| \gg 1/a$:

$$U(x) \approx \sqrt{\frac{2}{\sigma_{-2}}\left(a^2 - 2b\right)} \exp(-a|x|),$$
$$M(x) \approx \tilde{M}_0 + \tilde{M}_1 \exp\left(-\frac{|x|}{\sqrt{d}}\right), \qquad (19)$$

where the coefficients are

$$\tilde{M}_0 = \frac{1}{2}a^2 - b, \quad \tilde{M}_1 = \frac{1}{2\sqrt{d}}\int_{-\infty}^{+\infty}\cosh\left(\frac{x}{\sqrt{d}}\right)\left[\sigma(x)U^2(x) - \tilde{M}_0\right]dx, \qquad (20)$$

cf. exact solutions (8) and (13). In fact, the exponentially decaying term in the asymptotic expression (19) for $M(x)$ makes sense only for $4da^2 > 1$, i.e., if the correlation length of the nonlocality, $\sqrt{d}$, exceeds the modulation scale, $(2a)^{-1}$, otherwise the dominant exponentially decaying term switches to $\tilde{M}_2\exp(-2a|x|)$ with $\tilde{M}_2 = 4(1 + 2\sigma_0/\sigma_{-2})a^2(a^2 - 2b)(3a^2 - 16da^4 + 2b)^{-1}$, cf. Eqs. (8) and (13). Equations (19) and (20) demonstrate that the arbitrary propagation constant from the semi-infinite interval, $b < a^2/2$, determines the asymptotic form of the soliton solution, except for coefficient $\tilde{M}_1$, the integral expression for which includes the (generically) unknown function, $U(x)$. As concerns particular exact solutions (8) and (13), it is easy to check that $\tilde{M}_1$, as given by Eq. (20), *precisely* vanishes for them, therefore $\exp(-|x|/\sqrt{d})$ does not appear in these solutions.

The switch between the different asymptotic forms at $4da^2 < 1$ and $4da^2 > 1$, i.e., between the "less nonlocal" and "more nonlocal" systems, is a direct manifestation of the above-mentioned competition between the two different spatial scales. In that sense, exact solutions (8) and (13) are exceptional ones, which always demonstrate the "quasi-local" structure.

In fact, $b$ takes strictly negative values for solitons, on the contrary to models with the self-focusing nonlinearities, in which the bright-soliton's propagation constant is always positive. Indeed, if Eq. (4) is considered as the stationary linear Schrödinger equation with effective potential $M(x)$ in the form of a potential well, which has $M''(x=0) > 0$ [see Fig. 1(b) below], Eq. (5), taken at $x = 0$, yields $M(x=0) > 0$, hence $M(x) > 0$ holds at all $x$. Further, the soliton shape of $U(x)$ implies the existence of inflexion points, $x_{\text{infl}}$, at which $M'' = 0$. Finally, at these points Eq. (4) yields $b = -M(x_{\text{infl}}) < 0$.

In the quasi-local limit, which corresponds to $d \to 0$ at fixed $a$, Eq. (2) yields

$$m \approx \sigma(x)|u|^2, \qquad (21)$$

and Eq. (1) reduces to the known equation, $iu_z + (1/2)u_{xx} - \sigma(x)|u|^2 u = 0$, which gives rise to a completely stable family of fundamental solitons [17-20]. In fact,

Eq. (21) holds as the universal asymptotic relation at $|x|\to\infty$, with $m \approx \tilde{M}_0 = \text{const}$, see Eqs. (19) and (20) [except for the ultra-nonlocal limit, which corresponds to Eq. (22) presented below]. Then, the substitution of expression $|u(x)|^2 \approx \tilde{M}_0/\sigma(x)$, following from Eq. (21), into definition (6) of the total power leads to the above-mentioned conclusion: the system gives rise to self-trapped modes with the convergent total power if $\sigma(x)$ grows at $|x|\to\infty$ faster than $|x|$ (or faster than $r^2$, in the 2D version of the system).

In the opposite "ultra-nonlocal" limit, which corresponds to very large values of the squared correlation length, $d$, term $M$ in Eq. (5) may be neglected in comparison with $dM''$, reducing the equation to

$$M'' = -d^{-1}\sigma(x)U^2. \qquad (22)$$

Next, if Eq. (4) is considered, as above, as the stationary linear Schrödinger equation with potential $M(x)$, Eq. (22) gives rise to the effective potential which is convex at all $x$ [$M''(x) \leq 0$], hence this potential is *anti-trapping* [on the contrary the trapping potential produced by Eq. (5), see, e.g., the potential in Eq. (8) and in Fig. 1(b)], and solitons cannot exist in this case. This conclusion is supported by particular exact solutions (10) and (16), which demonstrate that the amplitudes of the solitons diverge in this limit.

All the above results, including asymptotic relations (19), (20) and the proof of $b < 0$, as well as the conclusion about the nonexistence of the solitons in the limit of $d \to \infty$, pertain equally well to DS modes, with a single obvious change that the asymptotic expression for $U(x)$ in Eq. (19) acquires an extra factor of sign($x$). In other words, the asymptotic form of the solitons at $|x|\to\infty$ does not depend on the structure of the solutions at smaller $|x|$.

## 4. Numerical results

As said above, numerical results for the family of fundamental solitons were obtained for the model based on Eqs. (1) and (2) with the nonlinearity-modulation function taken as per Eqs. (7), (9), and (18). By means of obvious rescaling, we fix $d=1$. Then, the degree of the nonlocality is controlled by $a$ in Eq. (7), the weakly and strongly nonlocal systems corresponding, respectively, to small and large values of $a$.

The simulations were performed by means of the split-step fast-Fourier-transform method. We have concluded that domain $-20 < x < +20$, in which the simulations were run, is sufficiently large to avoid effects of boundaries on soliton modes.

The stability of the stationary patterns was tested in direct simulations of Eq. (1), by adding a random noise to the initial conditions, with the strength amounting to 10% of the soliton's amplitude. In fact, this noise represents strong perturbations, hence persisting modes are truly robust ones.

### 4.1. Fundamental solitons in the single-channel system

The total power of the fundamental solitons, $P$ [see Eq. (6)], as obtained from the numerical solution of Eqs. (4) and (5), is displayed versus propagation constant $b$ in Fig. 1(a). As shown in the figure, this numerically found soliton family includes the particular exact solution, given by Eqs. (8), (10), and (11) [its profile is displayed in Fig. 1(b)].

Direct simulations have demonstrated that the entire fundamental-soliton family is stable. However, it was also observed that the solitons become, in a certain sense, more "fragile" with the increase of *a*, i.e., increase of the nonlocality degree. Namely, the simulated evolution of the perturbed soliton may seem unstable if the stepsize of the numerical scheme, $\Delta z$, is not small enough. For instance, the simulations for $a \sim 400$ produce a seemingly unstable evolution with $\Delta z = 10^{-5}$, but stable with $\Delta z = 10^{-6}$ (not shown here in detail). In fact, this observation can be explained by the fact that, according to Eq. (1), the respective spatial-modulation scale, $1/a$, corresponds to the characteristic diffraction length $z_{\text{diffr}} \sim 1/a^2 \sim 10^{-5}$, and the reliable simulations require, obviously, $\Delta z \ll z_{\text{diffr}}$.

**4.2. Fundamental solitons in the multi-channel system**

Next, we introduce the model with the *double-channel* modulation of the nonlinearity coefficient, which is defined as

$$\sigma(x) = \sigma_1(x - S/2) + \sigma_1(x + S/2) , \qquad (23)$$

where *S* is the distance between the channels, and $\sigma_1(x \mp S/2)$ is the single-channel modulation profile defined as per Eqs. (7), (9), and (18). In this case, the input for the simulations was taken as the linear superposition of the above solutions which are exact for the single-channel model [see Eqs. (8), (10), and (11)]:

$$U(x) = U_1(x - S/2) + U_2(x + S/2), M(x) = M_1(x - S/2) + M_2(x + S/2). \qquad (24)$$

It has been found that the double-channel configuration supports the stable propagation of the soliton pair if the individual solitons are stable in the respective single-channel setting, and separation $S$ is larger than a certain critical value, $S_{cr}$. An example of the ensuing soliton pair and its stable propagation is displayed in Figs. 1(c,e) for $S = 4$. At $S < S_{cr}$, the two-soliton propagation regime is subject to an oscillatory instability, which is caused by transverse interaction of the solitons trapped in the parallel channels, as shown in Fig. 1(f). The critical separation, $S_{cr}$, is shown as a function of the inverse with $a$ of the individual channels in Fig. 1(g). The decrease of $S_{cr}$ with the increase of $a$ is a natural feature, as smaller separation is sufficient to suppress the interaction between narrower solitons.

We have also checked properties of antisymmetric soliton pairs constructed, in the double-channel configuration, as $U(x-S/2) - U(x+S/2)$, cf. Eq. (24). It was concluded that their behavior is not essentially different from that of the symmetric pairs.

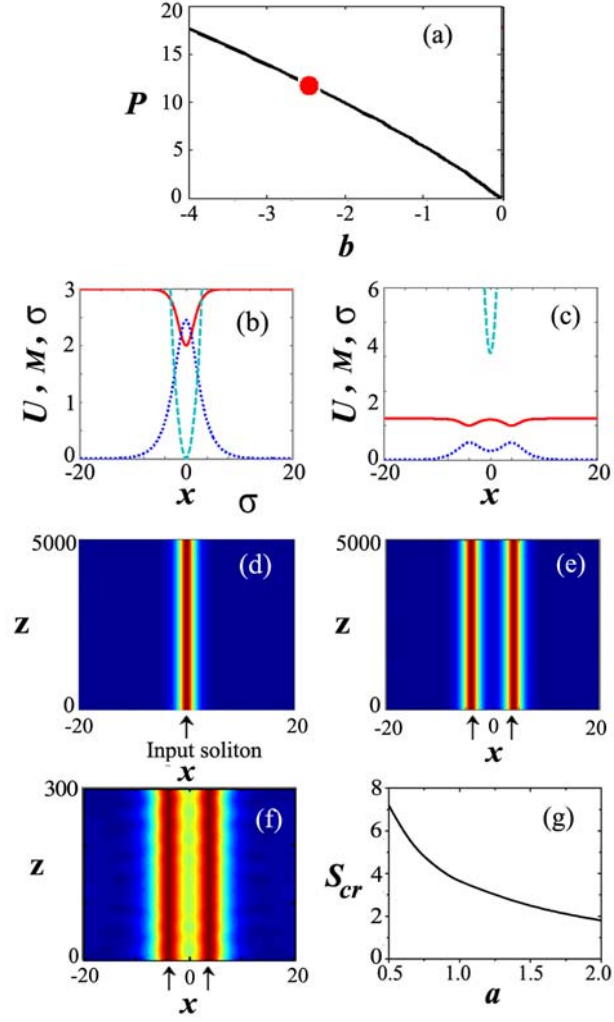

Fig. 1. (Color online) (a) The total power of the fundamental solitons versus the propagation constant, as produced by the numerical solution of Eqs. (4) and (5) with the nonlinearity-modulation function defined as per Eqs. (7), (9), and (18), with $d = 1$ and $\sigma_2 = 1$. The bold dot designates the value of $b$ for the exact solution, given by Eq. (11) with $a = 1$. (b,d) The shape and stable propagation of the soliton corresponding to exact solution [see Eqs. (8) and (10)] with $d = a = 1$ and $\sigma_2 = 1$. (c,e) The numerically produced shape and stable propagation of the pair of solitons in the double-channel modulation profile taken as in Eq. (22), with $S = 8$ and $a = 1$. In (b) and (c), the solid (red) curve represents the refractive-index perturbation, $M(x)$, the dotted (blue) curve shows the field shape, $U(x)$, and the dashed (magenta) curve is the profile of the nonlinearity-coefficient modulation, $\sigma(x)$. (f) An unstable soliton pair with $S = 8$ and $a = 0.5$. (g) The critical spacing between the adjacent channels, $S_{\mathrm{cr}}$, below which the propagation regime is unstable, versus the inverse-width parameter, $a$, of the individual channels, see Eq. (7). In panels (d), (e), and (f), and in similar figures displayed below, arrows indicate the input soliton. The same modulation profile and size of the spatial domain, $-20 \leq x \leq 20$, are adopted in all the figures displayed below.

For a similarly defined three-channel modulation profile,

$$\sigma(x) = \sigma_1(x - S) + \sigma_1(x) + \sigma_1(x + S) \quad , \tag{25}$$

Fig. 2(a) shows that the critical separation between the adjacent channels, necessary for the stable transmission of triple beams, is much larger than in the case of the double channel [cf. Fig. 1(a)]. Below the critical value, the attraction between the individual solitons gives rise to an oscillatory instability. Examples of stable and unstable three-soliton states are displayed in Figs. 2(b) and 2(c). It is seen that the oscillatory instability chiefly affects the soliton in the central channel.

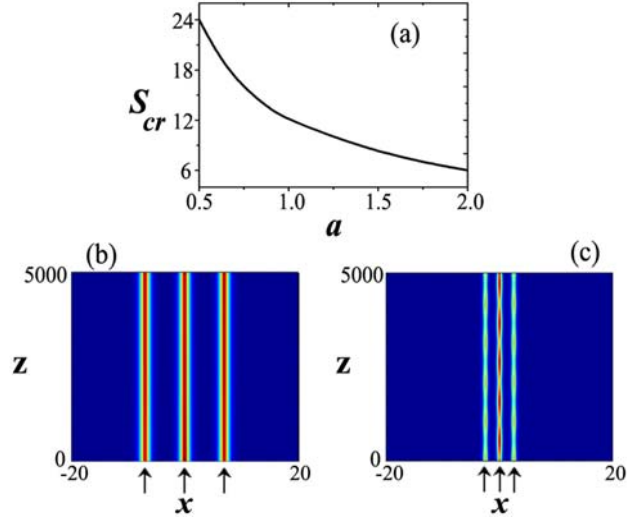

Fig. 2. (Color online) The three-soliton propagation in the triple waveguide defined as per Eq. (24). (a) The critical spacing between the adjacent channels, below which the propagation regime is unstable, versus parameter $a$. (b,c) Example of the stable and unstable propagation, for $S = 7$ and $a = 2$, and $S = 5$ and $a = 2$, respectively.

Next, we consider the propagation of a single soliton in the two-channel system, launching the beam into one channel, while the other one is left empty. It is found that when the inter-channel spacing, $S$, is small enough, the soliton periodically switches (jumps) between the channels, due to the penetration of the soliton's field into the parallel channel. This is an example of *Josephson oscillations* of the soliton in the effective nonlinear potential induced by the spatial modulation of the nonlinearity strength.

Depending on the inverse channel's width, $a$, there is a largest (critical) value of $S$ above which the soliton does not perform the Josephson oscillations, as shown in Fig. 3(a). Examples of the Josephson-oscillation regime for a single soliton in the two- and three-channel systems, which correspond to the modulation profiles defined by Eqs. (23) and (25), are shown in Fig. 3(b) and Figs. 3(c,d), respectively. In Fig. 3(b), the soliton keeps its shape in the course of the oscillations between the two channels, while in Figs. 3(c) and 3(d) the soliton's power becomes smaller in the middle core of the three-channel system.

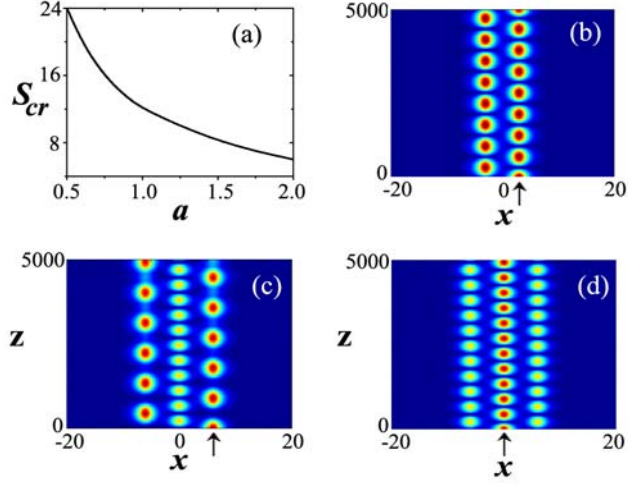

Fig. 3. (Color online) The Josephson regime for a single soliton periodically switching between adjacent cores in the two- and three-channel systems. (a) The critical spacing between the adjacent waveguides versus the inverse width, $a$ [see Eq. (7)], for the double-channel waveguide. Below the critical value, the solitons switch between adjacent waveguides, while above it they stay in the original waveguide. (b) The original single soliton periodically switching between two channels, with $S = 6$ and $a = 1$. (c) The soliton originally launched into the edge core in the three-channel system is periodically switching among the adjacent ones, with $S = 6$ and $a = 1$. (d) The soliton originally launched into the central core of the three-channel system is periodically switching between the three channels, with $S = 6$ and $a = 1$.

Next, we consider two-soliton propagations in the three-channel setting. In this case, the two solitons can again jump (switch) between the adjacent channels, provided that spacing $S$ is not too large. The corresponding critical (largest) value of $S$ admitting the periodic switching is plotted versus the inverse channel's width, $a$, in Fig. 4(a). This critical value is obviously larger than its counterpart in Fig. 3(a), because the inter-channel overlap, which is responsible for switching, is produced by two solitons. Examples of the two-soliton Josephson-oscillation regimes in the three-channel system are shown in Fig. 4(b) and Figs. 4(c).

In Fig. 4(b), the power of the solitons launched into the right and left channels is partly transferred into to the middle one, which makes the soliton's power larger in the middle channel. Then the power flows back to the edge channels, and this coupling regime repeats itself periodically. In Fig. 4(d), the solitons, initially launched into the right and middle channels, do not couple to adjacent ones, as the spacing between them is too large. To the best of our knowledge, Josephson oscillations of solitons were not reported before in other nonlocal models.

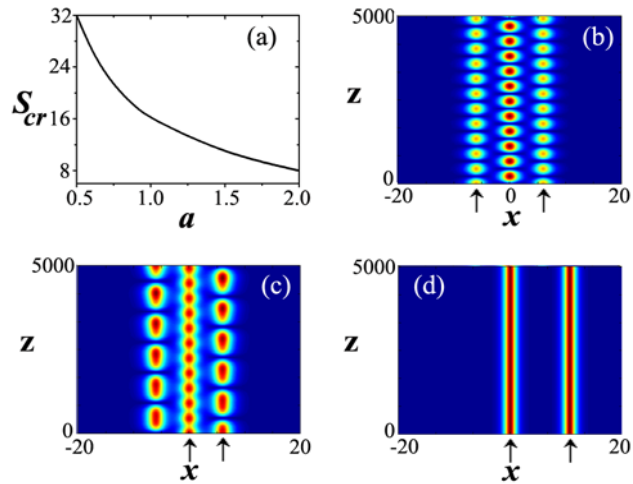

Fig. 4. (Color online) The two-soliton propagation in the three-channel system. (a) The critical spacing between the adjacent channels versus $a$, which has the same

meaning as in Fig. 3(a). (b) Two solitons, launched into the edge channels, with $S = 6$ and $a = 1$, periodically switch between the three channels. (c) The same for two solitons launched into the right and central channels. (d) The solitons launched into two channels cannot couple into adjacent ones, as the spacing between them is too large, $S = 16$, with $a = 1$.

Dynamical regimes in the multichannel settings, similar to those reported above, may be expected as well in the local counterpart of the model, which corresponds to $d \to 0$ (multichannel configurations were not considered previously in the local models with the self-trapping provided by the local self-defocusing strength growing from the center to periphery [17-21]). However, our analysis has demonstrated that Josephson oscillations of solitons between the channels are *always unstable* in the local version of the present system (note shown here in detail). This finding may be explained by the fact that the nonlocality, which additionally couples adjacent channels, is necessary for the stabilization of the periodic jumps of the solitons between the channels.

### 4.3. Dipole solitons (DSs)

Generic results for families of DS solutions can be presented for the modulation profile given by Eqs. (14) and (15) with $\sigma_0 = 1$ [and fixing, as said above, $d = 1$ in Eq. (2)]. First, in Fig. 5(a) we display the total power [see Eq. (6)] of the numerically found DS family versus the propagation constant, which includes the exact solution given by Eqs. (16)-(17) [see panel 5(b)]. We have checked that the entire DS family is completely stable, see an example shown in Figs. 5(b) and 5(d).

The multi-DS propagation in the multi-channel systems was studied too. The character of the propagation again depends on spacing $S$ between adjacent channels. If $S$ is smaller than a respective critical value, $S_{\mathrm{cr}}$, the DS complexes are unstable, as the interaction between the DSs trapped in adjacent channels is too strong. Examples of the stable and unstable multi-DS propagation are displayed in Figs. 5(c,e) and 5(f), respectively. The corresponding value of $S_{\mathrm{cr}}$ is displayed, as a function of $a$ [the inverse-width parameter in Eq. (14)], in Fig. 5(g).

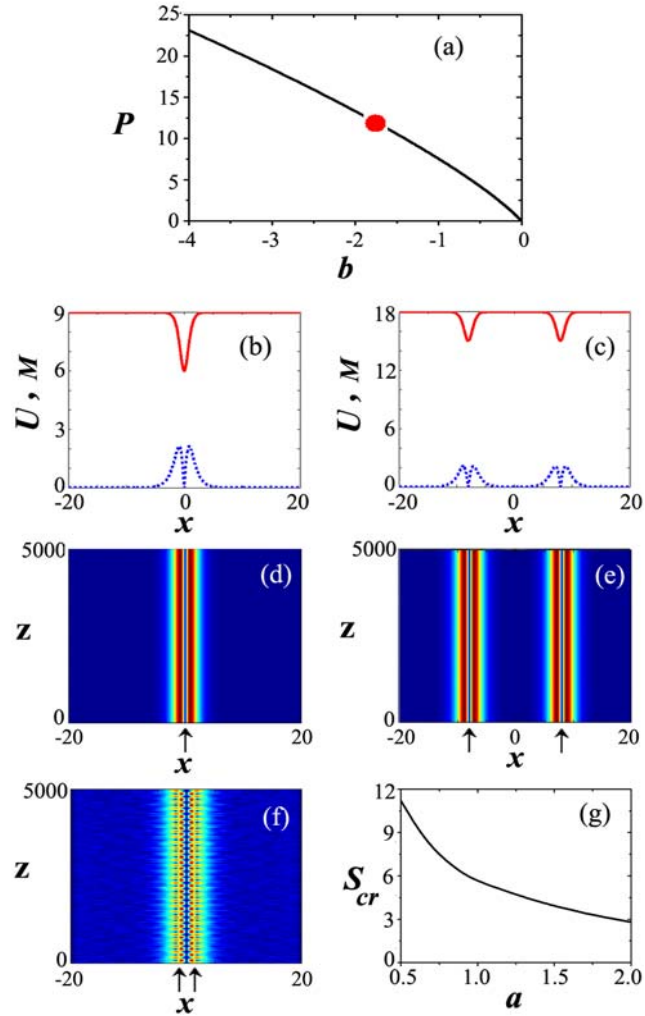

Fig. 5. (Color online) The propagation of single dipole solitons (DSs) and DS pairs in the respective waveguides. (a) The total power of the numerically found DS family versus the propagation constant. The bold dot designates the value of *b* for the exact solution, as given by Eq. (17) with $d = a = 1$. (b,d) The profile and evolution of the stable exact DS solution. (c,e) A stable DS pair, with large spacing $S = 16$. In (b) and (c), the solid (red) curves represent the refractive-index perturbation, $M(x)$, and the dotted (blue) curves represent the field shape, $U(x)$. (f) Unstable co-propagation of two DSs in the two-channel system, when the spacing is too small, $S = 2$. (g) The critical spacing, $S_{cr}$, versus *a* [see Eq. (14)], the propagation of the DS pair being unstable at $S < S_{cr}$.

Finally, we consider the Josephson-oscillation regimes for the DS transmission in two- and three-channel systems. If the spacing between the adjacent channels is neither too large nor too small, the DS can stably switch between them, similar to the Josephson dynamical regime reported above for the fundamental solitons, cf. Fig. 3. However, when the spacing is smaller than another critical value, the DSs exhibits *unstable oscillations* between the channels (a regime which was not observed for the fundamental solitons). The corresponding critical values are displayed in Fig. 6(a). Figures 6(b)-6(d) represent the stable Josephson regime for the single DS. Figures 6(e) and 6(f) show an example of a pair of DSs stably switching among three channels. An example of unstable jumps of the single DS between two channels is shown in Fig. 6(g), and unstable jumps of a DSs pair between three channels are displayed in Fig. 6(h), for a small spacing between the adjacent channels. In the latter case, conspicuous emission of radiation takes place due to strong interaction between the DSs.

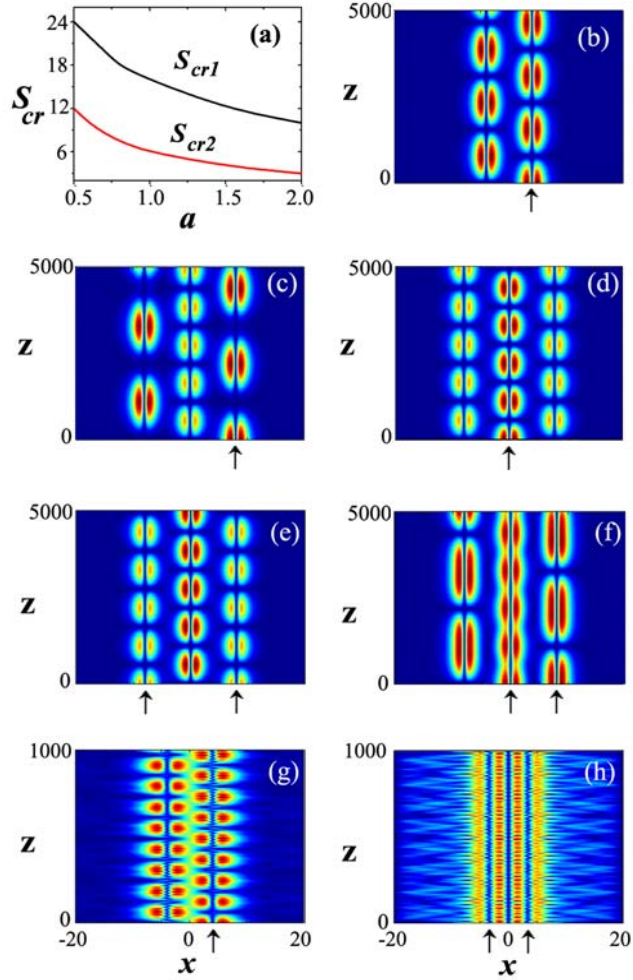

Fig. 6. (Color online) Periodic switching of DSs between adjacent channels. (a) The critical spacing between adjacent waveguides, $S_{cr}$, versus $a$. At $S_{cr1} < S < S_{cr2}$, DSs feature stable Josephson oscillations between adjacent waveguides, while at $S < S_{cr1}$ DSs perform unstable oscillations and at $S > S_{cr2}$ DSs stay at original waveguides and stably propagate. (b) The regime of stable periodic oscillations of a single DS between two adjacent channels, with $S = 8$ and $a = 1$. (c) Stable oscillations of a single DS in the three-channel system with $S = 8$ and $a = 1$. (d) Stable oscillations of a single DS launched into the central core of the three-channel system, with $S = 8$ and $a = 1$. (e) Stable oscillations of two DSs launched into the edge cores of the three-channel system, with $S = 8$ and $a = 1$. (f) Stable oscillations of two

DSs launched into the right and central cores of the three-channel system, with $S = 8$ and $a = 1$. (g) and (h) Unstable oscillations of a single DS between two and three channels, with $S = 4$ and $a = 0.5$.

## 5. Conclusion

We have introduced the 1D model of an optical medium with the strength of the nonlocal self-repulsive nonlinearity growing from the center to periphery. The model can be derived for the propagation of light in a slab waveguide with the thermal nonlinearity, in which the growing nonlinearity is provided by the respective profile of the concentration of absorptive dopants which heat the material. Following the recently elaborated mechanism of self-trapping of stable localized modes in local media with the self-defocusing nonlinearity whose strength grows at $|x|\to \infty$ faster than $|x|$, we have found stable fundamental and dipole-mode solitons in the nonlocal model. The essential difference from the local media is the competition between the two different spatial scales, imposed by the spatial modulation of the nonlinearity coefficient, and by the nonlocality. The competition explicitly manifests itself in the analytically found asymptotic form of the generic solutions. Particular exact solutions were produced too, for both the fundamental and dipole solitons. Interestingly, the exact solutions, unlike the generic ones, effectively evade the competition between the two spatial scales, always featuring the structure which resembles solutions of the local system. Multi-channel systems, with two or three local minima of the nonlinearity coefficient, have been considered too (such settings were not studied previously in local counterparts of the present model). They support stable propagation of several solitons along the channels, provided that the spacing between them exceeds a certain critical value. In the Josephson regime, the solitons and dipoles periodically switch between the channels. A noteworthy finding is that stable Josephson regimes are not possible in the local version of the present system.

These results suggest new possibilities for experimental and theoretical studies of the solitons dynamics in nonlocal media with various modulation patterns. It may be interesting to extend the analysis for similar two-dimensional settings, which will, however, require a more careful physical justification. Another challenging direction for further studies is to consider similar possibilities in Bose-Einstein condensates with long-range interactions [22].

**Appendix: the derivation of the model for the slab waveguide**

Equation (2) can be derived as the heat-conductivity equation, with the source term on the right-hand side, for the scaled temperature disturbance, $m(x,z)$, which gives rise to the local change of the refractive index (as usual, it is assumed that the time derivative in the equation may be omitted [1]). To this end, we consider a slab waveguide, with the transverse and longitudinal coordinates, $x$ and $z$, in the plane of the slab, and perpendicular coordinate $y$. The corresponding stationary heat-conductivity equation amounts to the 2D Poisson equation:

$$m_{xx} + m_{yy} = -\sigma(x,y)|u(x,y,z)|^2, \tag{A1}$$

where it is assumed that term $\partial^2 m/\partial z^2$ in the Poisson equation may be neglected, as the variation of the temperature-disturbance field is much faster along $x$ and $y$ than along the propagation distance, $z$. Further, the transverse structure of the light-field amplitude in the slab may be naturally approximated by

$$u(x,y,z) = Y(y)u(x,y), \quad (A2)$$

where $Y(y)$ is the slab's transverse modal function [its exact shape is not crucially important, as it only determines one constant, see Eq. (A5) below].

The thickness of the slab (along $y$) being $h$, the simplest distribution of the heat-inducing dopant is

$$\sigma(x,y) = \begin{cases} \cos(\pi y/h)\widetilde{\sigma}(x) & \text{at } |y| < h/2, \\ 0 & \text{at } |y| \geq h/2. \end{cases} \quad (A3)$$

Accordingly, the solution to Eq. (A1) inside the slab may be sought for as $m(x,y) = m(x)\cos(\pi y/h)$, where $m(x)$ satisfies the following inhomogeneous equation, which follows from Eqs. (A1) and (A2):

$$m'' - (\pi/h)^2 m = -Y_0^2 \widetilde{\sigma}(x)|u(x,z)|^2, \quad (A4)$$

and the modal coefficient,

$$Y_0^2 \equiv \frac{2}{h}\int_{-h/2}^{+h/2}\left[\cos^2\left(\frac{\pi y}{h}\right)\right]Y^2(y)dy, \quad (A5)$$

is determined by the projection of the right-hand side of Eq. (A1) onto the transverse function $\cos(\pi y/h)$ from Eq. (A3). Finally, Eq. (A4) is tantamount to Eq. (2) with $d \equiv (h/\pi)^2$ and $\sigma(x) \equiv Y_0^2(\pi/h)^2 \widetilde{\sigma}(x)$. In particular, this setting give rise to a natural value of the correlation length, determined by the slab's thickness, $\sqrt{d} = h/\pi$.

**Acknowledgments**

We appreciate a valuable discussion with Y.V. Kartashov. This work was supported by the National Natural Science Foundation of China (Grant No. 11174061) and the Guangdong Province Natural Science Foundation of China (Grant No. S2011010005471).